\documentclass[pra,twocolumn,superscriptaddress]{revtex4}
\usepackage[utf8]{inputenc}
\usepackage{amsmath}
\usepackage{amssymb}
\usepackage{graphicx}
\usepackage{esint}
\usepackage{subfigure}
\usepackage[colorlinks,citecolor=blue]{hyperref}

\makeatletter

\@ifundefined{textcolor}{}
{%
 \definecolor{BLACK}{gray}{0}
 \definecolor{WHITE}{gray}{1}
 \definecolor{RED}{rgb}{1,0,0}
 \definecolor{GREEN}{rgb}{0,1,0}
 \definecolor{BLUE}{rgb}{0,0,1}
 \definecolor{CYAN}{cmyk}{1,0,0,0}
 \definecolor{MAGENTA}{cmyk}{0,1,0,0}
 \definecolor{YELLOW}{cmyk}{0,0,1,0}
}

\usepackage{subfigure}\usepackage{epsfig}\usepackage{amsfonts}\usepackage{mathrsfs}\usepackage{CJK}\usepackage[toc,page,title,titletoc,header]{appendix}

\makeatother

\begin{document}

\title{Enhancing Kondo Coupling in Alkaline-Earth Atomic Gases with Confinement-induced Resonances in Mixed Dimensions}
\author{Yanting Cheng}
\affiliation{Institute for Advanced Study, Tsinghua University, Beijing, 100084, China}
\author{Ren Zhang}
\email{rine.zhang@gmail.com}
\affiliation{Institute for Advanced Study, Tsinghua University, Beijing, 100084, China}
\author{Peng Zhang}
\email{pengzhang@ruc.edu.cn}
\affiliation{Department of Physics, Renmin University of China, Beijing, 100872,
China}
\affiliation{Beijing Computational Science Research Center, Beijing, 100084, China}
\author{Hui Zhai}
\affiliation{Institute for Advanced Study, Tsinghua University, Beijing, 100084, China}
\affiliation{Collaborative Innovation Center of Quantum Matter, Beijing, 100084, China}

\date{\today}

\begin{abstract}
The Kondo effect describes the spin-exchanging interaction between localized impurity and the itinerant fermions. The ultracold alkaline-earth atomic gas provides a natural platform for quantum simulation of the Kondo model, utilizing its long-lived clock state and the nuclear-spin exchanging interaction between the clock state and the ground state. One of the key issue now is whether the Kondo temperature can be high enough to be reached in current experiment, for which we have proposed using a transverse confinement to confine atoms into a one-dimensional tube and to utilize the confinement-induced resonance to enhance the Kondo coupling. In this work, we further consider the $1+0$ dimensional scattering problem when the clock state is further confined by an axial harmonic confinement. We show that this axial confinement for the clock state atoms not only plays a role for localizing them, but also can act as an additional control knob to reach the confinement-induced resonance. We show that by combining both the transverse and the axial confinements, the confinement-induced resonance can be reached in the practical conditions and the Kondo effect can be attainable in this system.  
\end{abstract}

\maketitle

\section{Motivation and Background}

In the past decades, experiments in cold atom systems have successfully explored many intriguing quantum many-body phenomena of different paradigms, including fermion pairing and the BCS-BEC crossover, the Bose and the Fermi Hubbard models \cite{bloch-review}, the Kosterlize-Thouless transition \cite{bloch-review}, one-dimensional integrable models \cite{guan}, spin-orbit coupling \cite{zhai-soc} and topological models \cite{topo}. Exploring these phenomena with cold atom systems have a list of advantages, for instance, one can access physical quantities that have not been measured before in their condensed matter realizations, and one can also study non-equilibrium dynamics in a highly controllable way. However, until now there is still one important category that has not been experimentally realized with cold atom systems yet, despite of quite a few existing proposals \cite{stoof,duan,rey,carmi,demler,nishida,kawakami,rey2,ren-kondo,kikoin}, and that is the Kondo physics.

\begin{figure}[t]
\begin{center}
\includegraphics[width=2.in]{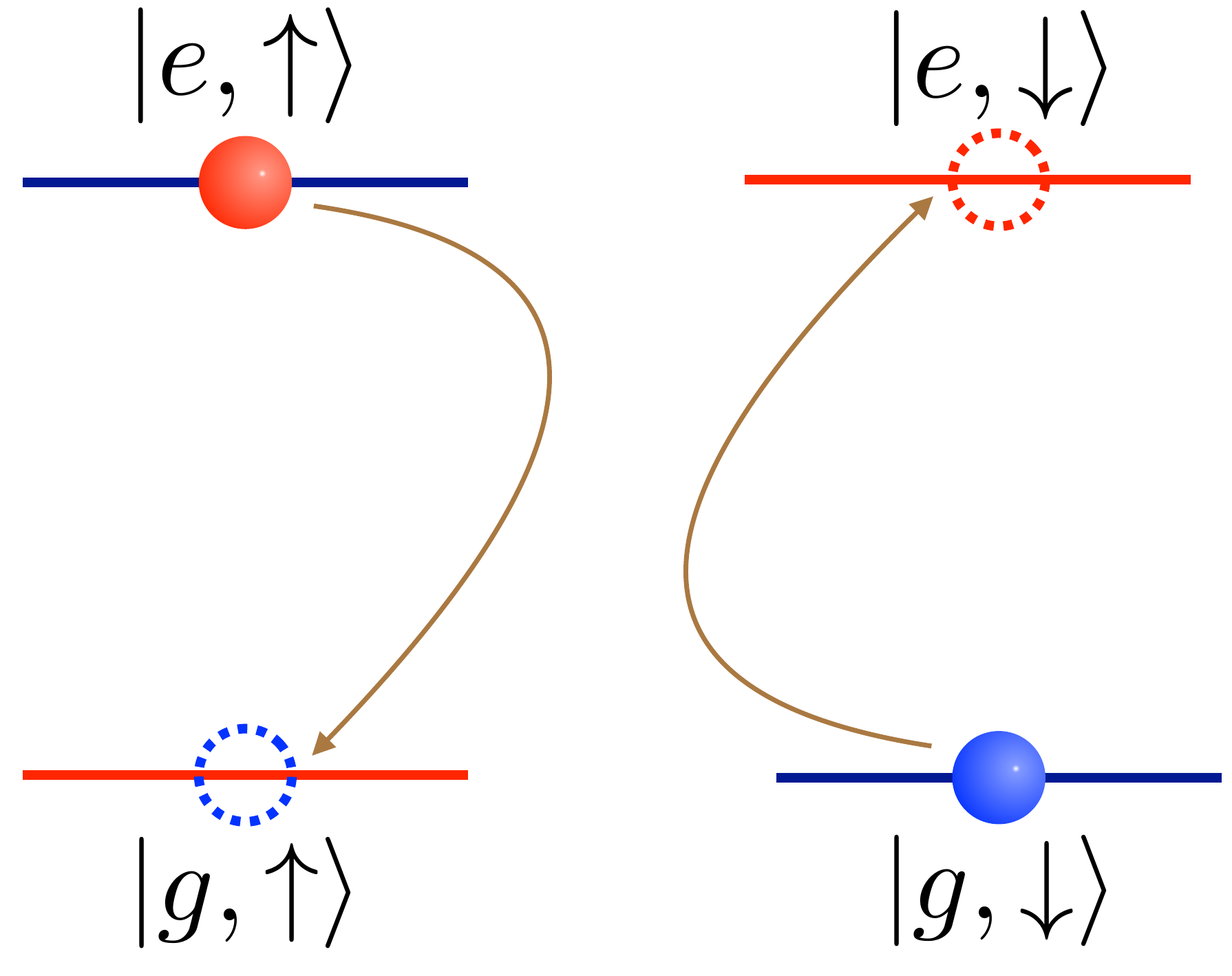}
\end{center}
\caption{Schematic energy level for alkaline-earth atoms. The red ball and red dashed circle represent $|e\rangle$-atom(long-lived clock state). The blue ball and blue dashed circle represent $|g\rangle$-atom(ground state). Nuclear spin-exchanging happens when $|e\rangle$-atom and $|g\rangle$-atom collide.
\label{schematic0}}
\end{figure}

\begin{figure}[t]
\begin{center}
\includegraphics[width=3.in]{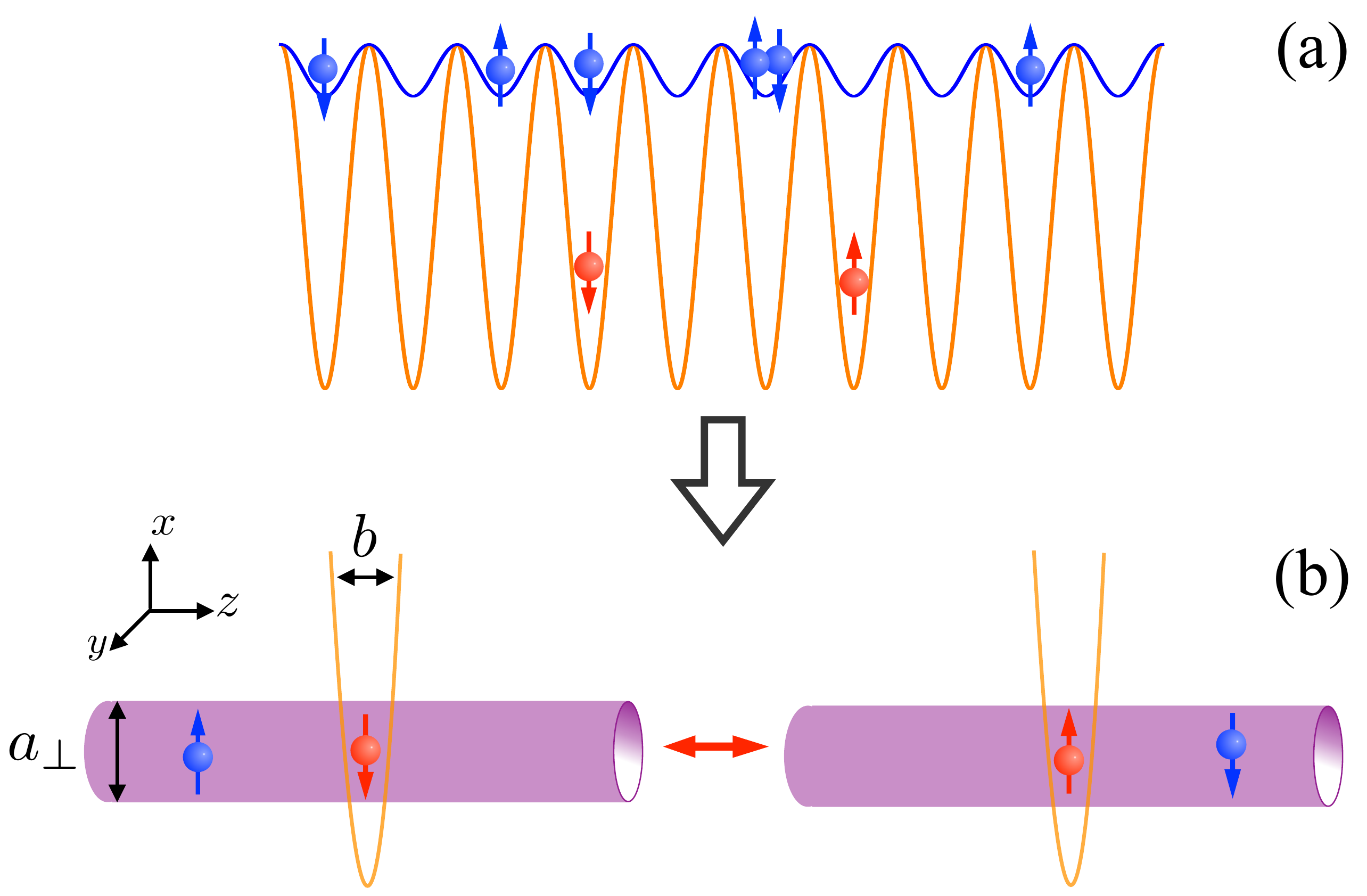}
\end{center}
\caption{(a): A pair of laser beams creates a deep lattice for atoms in the clock state $|e\rangle$ and a weak lattice for atoms in the ground state $|g\rangle$.  (b): The key process responsible for the Kondo effect is the spin-exchanging scattering between impurity (localized $|e\rangle$-atom) and the itinerant fermions ($|g\rangle$-atoms). The blue and red ball denote the ground state and the clock state, respectively. The nuclear spins are denoted by arrows.
\label{schematic}}
\end{figure}

The Kondo model describes the spin-exchanging interaction between localized impurities and the itinerant fermions \cite{Kondo}. The alkaline-earth atomic gases have natural advantages for performing quantum simulation of the Kondo model. The schematic energy level of single alkaline-earth atoms is shown in Fig.~\ref{schematic0}. First of all, there is a long-lived electronic excited state known as the clock state, usually denoted by $|e\rangle$.  Atoms in this clock state generically has a different ac polarization comparing to atoms in their electronic ground state, usually denoted by $|g\rangle$, except for lasers with a magic wavelength \cite{magic_wave1,magic_wave2}. Therefore, it is easy to realize a situation that lasers create a deep lattice for $|e\rangle$-atoms and make them localized as impurities, while $|g\rangle$-atoms experience a quite shallow lattice and remain itinerant, as shown in Fig. \ref{schematic}(a). 

Secondly, in addition to the orbital degree of freedom labelling $|g\rangle$ and $|e\rangle$, there is also a nuclear spin degree of freedom. Due to the nearly perfect decoupling between the nuclear spin and the electronic degree of freedoms in these two states of alkaline-earth atoms \cite{Jun_PRA}, the interaction has to be invariant under the nuclear spin rotation. Thus, the interaction has to be diagonal in the nuclear spin singlet and three triplet channels, and for the $s$-wave interaction, due to the symmetrization condition of the entire wave function, these channels also have to be orbital triplet and singlet, respectively. Considering the interaction between one atom in the $|g\rangle$-state and one atom in the $|e\rangle$-state, the relevant bases are \cite{Munich-spin-exchange,Florence-spin-exchange}
\begin{align}
&|+\rangle=\frac{1}{2}(|ge\rangle+|eg\rangle)\left(|\uparrow\downarrow\rangle-|\downarrow\uparrow\rangle\right),\\
&|-,0\rangle=\frac{1}{2}(|ge\rangle-|eg\rangle)\left(|\uparrow\downarrow\rangle+|\downarrow\uparrow\rangle\right),\\
&|-,1\rangle=\frac{1}{\sqrt{2}}(|ge\rangle-|eg\rangle)|\uparrow\uparrow\rangle,\\
&|-,-1\rangle=\frac{1}{\sqrt{2}}(|ge\rangle-|eg\rangle)|\downarrow\downarrow\rangle,
\end{align}
where $+$ and $-$ denote orbital triplet and singlet, respectively, and in $|-,q\rangle$, $q$ denotes total nuclear spin along the $\hat{z}$ direction.

In three dimension, the two-body interaction can be written as
\begin{equation}
V({\bf r})=V_{+}({\bf r})|+\rangle\langle +|+\sum\limits_{q=0,\pm 1}V_{-}({\bf r})|-,q\rangle \langle -,q|. \label{int}
\end{equation}
where ${\bf r}$ is the relative coordinate between a $|g\rangle$- and an $|e\rangle$-atom. 
$V_{+}$ and $V_{-}$ are both short-ranged potentials, and they are described by two scattering lengthes of $a_{s,+}$ and $a_{s,-}$, respectively. Expanding $V({\bf r})$ into the bases of 
\begin{align}
|g\uparrow; e\downarrow\rangle=\frac{1}{\sqrt{2}}(|+\rangle+|-,0\rangle),\\
|g\downarrow; e\uparrow\rangle=\frac{1}{\sqrt{2}}(|-,0\rangle-|+\rangle),
\end{align}
and $|-,\pm 1\rangle$, 
the interaction becomes 
\begin{align}
\label{eqn:potential}
&V({\bf r})\nonumber\\
&=\frac{V_{+}({\bf r})+V_{-}({\bf r})}{2} (|g\uparrow; e\downarrow\rangle \langle g\uparrow; e\downarrow|
+|g\downarrow; e\uparrow\rangle \langle g\downarrow; e\uparrow|)\nonumber\\
&+\frac{V_{-}({\bf r})-V_{+}({\bf r})}{2} (|g\uparrow; e\downarrow\rangle \langle g\downarrow; e\uparrow|
+|g\downarrow; e\uparrow\rangle \langle g\uparrow; e\downarrow|)\nonumber\\
&+V_{-}({\bf r})|-,1\rangle \langle -,1|+V_{-}({\bf r})|-,-1\rangle \langle -,-1|.
\end{align}
Here one can see that the key point is that the difference of between $V_{+}$ and $V_{-}$ gives rise to a spin-exchanging interaction between $|g\rangle$-atoms and $|e\rangle$-atoms, as shown in Fig. \ref{schematic}(b), whose effect is the most profound in the zero-field limit when the single particle energies of $|g\uparrow; e\downarrow\rangle$ and $|g\downarrow; e\uparrow\rangle$ are degenerate. Actually, such spin-exchanging processes have been observed in experiments \cite{Munich-spin-exchange,Florence-spin-exchange}. In this magnetic field regime, when $|g\rangle$-atoms are itinerant and $|e\rangle$-atoms are localized, this will give rise to a Kondo model. 

It looks like that everything is ready. However, there is a last challenging to overcome. That is, since in cold atom system, it is hard to achieve the temperature regime that is orders of magnitude lower than the Fermi temperature, it is important to increase the strength of the spin-exchanging interaction such that the Kondo temperature becomes high enough to be attainable by current cooling capability. 

To solve this problem, in the previous work, we propose a scheme to use confinement-induced resonance to increase the Kondo coupling \cite{ren-kondo}. 
The basic idea is to use lasers with magic wave length to create a two-dimensional lattice, with which atoms in both $|g\rangle$-state and $|e\rangle$-state are confined into a one-dimensional tube with the same transverse confinement radius $a_\perp$. With the standard formula of the confinement-induced resonance \cite{CIR}, at zero-field the reduced interaction in one-dimensional still takes the form as Eq.~(\ref{int}), with $V_{\xi}=g_{\xi}\delta(z)$ ($\xi=+,-$), and 
\begin{align}
g_{\xi}=\frac{4\hbar^{2}a_{s,\xi}}{ma_{\perp}^{2}}\left(1-{\cal C}\frac{a_{s,\xi}}{a_{\perp}}\right)^{-1}, \quad {\cal C}\approx1.46035\cdots. \label{CIR}
\end{align}
Thus, when $a_\perp$ approaches either $\mathcal{C}a_{s,+}$ or $\mathcal{C}a_{s,-}$, one of the $g_{\xi}$ will diverge while the other remains finite. Hence, their difference will become very large and the Kondo effect will be enhanced. In Ref.~\cite{ren-kondo} we have also considered the correction due to finite Zeeman field, and we find that the effect is insignificant in the low-field regime. Hence, in this paper we only consider the zero-field case for simplicity.

Nevertheless, it still does not complete the whole story for two reasons. First, on the experimental side, in order for the system to be in the one-dimensional regime, $a_\perp$ can not be too large; and on the other hand, due to the practical constraint of the laser power, $a_\perp$ also can not be arbitrarily small. That is to say, there is a range to tune $a_\perp$, and within this range, maybe $a_\perp$ neither reach $\mathcal{C}a_{s,+}$ nor $\mathcal{C}a_{s,-}$. Second, on the theory side, in Ref. \cite{ren-kondo} the axial confinement for the localized $e$-atoms is only treated at the level of the lowest Wannier wave function approximation. Such an approximate may fail in certain parameter regime.  

Motivated by the experimental efforts along this direction \cite{Simon}, the present work treats a $1+0$ dimensional scattering problem which solves these two problems at once. Here we treat the axial confinement for $|e\rangle$-atoms beyond the Wannier wave function approximation, and we will show that it indeed gives rise to much richer features. In particular, we show that this axial confinement length $b$ (as shown in Fig. \ref{schematic}(b)) plays a role as an extra control parameter to reach the confinement-induced resonance. That is to say, even if $a_\perp$ is not very close to $\mathcal{C}a_{s,+}$ or $\mathcal{C}a_{s,-}$, one can still reach resonance by tuning $b$ with the axial confinement. 

\section{Model}

To be numerically tractable, we simplify the problem from the lattice case as shown in Fig. \ref{schematic}(a) to a 1+0 dimensional problem as shown in Fig. \ref{schematic}(b). This simplification includes following assumptions: i) For $|e\rangle$-atoms, we assume that it is always localized in one of the lattice site, and we expand the potential to the quadratic order nearby its minimum. That is to say, we ignore the anharmonic part of the lattice potential. The axial confinement problem is treated as an harmonic trap with frequency $\omega_z$.  ii) The laser beam still induces a weak lattice potential for the $|g\rangle$-atom, though the ac polarizations of $|g\rangle$-state and $|e\rangle$-state can differ a lot. Here we take this effect into account by an effective mass approximation, that to say, we still assume the single particle dispersion for $|g\rangle$-atom is a parabola but with an effective mass denoted by $m_g$. In general $m_g>m$ ($m$ is the bare mass). Since in one-dimension, the density-of-state diverges at zero-energy and the low-energy states contribute most to the scattering problem, we believe that the effective mass approximation is reasonably good in one-dimension. 

We will show that even with these two approximations, the results should be qualitatively good, and the results can be systematically improved later by including the anharmonicity for the localized atoms and other lattice effects for the itinerant atoms. Furthermore, in principle, we should solve the problem in three dimension by including the transverse confinement for both states and the axial confinement solely for $|e\rangle$-atoms in the equal footing, however, here simplify the procedure by first treating the transverse confinement with Eq.~(\ref{CIR}), and subsequently studying the $1+0$ dimension. We believe this treatment is sufficiently good because the resonances induced by transverse confinement and the axial confinement are two different physical processes and they generally will not strongly interfere with each other. 

We now start from the one-dimensional Hamiltonian as the form    
 \begin{align}
\hat{H}_{\rm tot}=\hat{H}_{+}|+\rangle\langle +|+\sum\limits_{q=0,\pm 1}\hat{H}_{-}|-,q\rangle\langle -,q|. \label{H1d}
\end{align}
Here
\begin{align}
\label{hamilton}
\hat{H}_{\xi}=-\frac{\hbar^{2}}{2m_g}\frac{d^{2}}{dz_{g}^{2}}-\frac{\hbar^{2}}{2m}\frac{d^{2}}{dz_{e}^{2}}+\frac{1}{2}m\omega^{2}_{z}z_{e}^{2}+g_{\xi}\delta(z_{e}-z_{g}),
\end{align}
for $\xi=+,-$, with $z_{g}$ and $z_{e}$ being the positions of $|g\rangle$-state and $|e\rangle$-state, respectively.

\section{Method}

Previously, scattering problem in the mixed dimension such as 3+0, 3+1 or 3+2 have been studied theoretically \cite{yvan-mixd-d,tan-mixd-d,tan2,zhigang} and experimentally \cite{mixd-exp}. Here the problem we considered is a $1+0$-dimension one. Since the Hamiltonian is diagonal in the bases chosen in Eq.~(\ref{H1d}), we only need to solve the Hamiltonian Eq.~(\ref{hamilton}) and the difference in the effective interaction strength between the $|+\rangle$ and $|-,0\rangle$ channels gives rise to the spin-exchanging strength.

Because of the additional harmonic trap along the axial direction for the $|e\rangle$-atom, the relative motion of the $|g\rangle$-atom and $|e\rangle$-atom is not separable from their center-of-mass motion. 
Here we consider the scattering between a moving $|g\rangle$-atom with momentum $\hbar k$ ($k>0$) along the $z$-direction and an $|e\rangle$-atom in the ground state of the trap.
The incident wave function is given by
\begin{align}
\psi_{k}^{0}(z_{g},z_{e})=\frac{1}{\sqrt{2\pi}}{\rm e}^{ikz_g}\phi_{0}(z_{e}),
\end{align}
where $\phi_{0}(z_{e})=\exp(-z_{e}^{2}/2b^{2})/\sqrt{b\pi^{1/2}}$ is the ground-state wave function of the harmonic trap. $b=\sqrt{\hbar/m\omega_{z}}$ is the harmonic trap length for the axial trap. For our system the two-body scattering wave function $\psi_{k}^{+}(z_{g},z_{e})$ is given by 
the Lippmann-Schwinger equation
\begin{align}
\label{eqn:scattering-state}
\psi_{k}^{+}(z_{g},z_{e})=\psi_{k}^{0}(z_{g},z_{e})+&g_{\xi}\int dz'G_{0}(z_{g},z_{e};z',z')\times\nonumber\\
&\hspace{1.5cm}\psi_{k}^{+}(z',z').
\end{align}
Here $G_{0}(z_{g},z_{e};z_{g}',z_{e}')$ is the two-body free Green's function. 
In this paper we consider the low-energy scattering process where $\hbar^2k^2/(2m_g)<\hbar\omega_z$.
For this case we have
\begin{align}
\label{eqn:green-function}
G_{0}(z_{g},z_{e};z_{g}',z_{e}')=&-i\frac{m_g}{\hbar^{2}}\frac{{\rm e}^{ik|z_{g}-z_{g}'|}}{k}\phi_{0}(z_{e})\phi_{0}^{*}(z_{e}')\nonumber\\
&\hspace{-2.2cm}-\frac{m_g}{\hbar^{2}}\sum_{n=1}^{\infty}\frac{{\rm e}^{-\sqrt{2m_gn\omega_z/\hbar^{2}-k^{2}}|z_{g}-z_{g}'|}}{\sqrt{2m_gn\omega_z/\hbar^{2}-k^{2}}}\phi_{n}(z_{e})\phi_{n}^{*}(z_{e}'),
\end{align}
with $\phi_{n}(z)=\sqrt{1/(b\sqrt{\pi}2^{n}n!)}\exp(-z^{2}/2b^{2})H_{n}(z/b)$ ($n=1,2,\cdots$) being the $n$th eigen wave function of the harmonic trap. 

Eqs.~(\ref{eqn:scattering-state}) and (\ref{eqn:green-function}) imply that in the limit $|z_{g}|\to\infty$ the 
scattering wave function $\psi_{k}^{+}(z_{g},z_{e})$ can be expressed as
\begin{eqnarray}
\psi_{k}^{+}(|z_g|\to\infty,z_e)& = & \frac{1}{\sqrt{2\pi}}\left[{\rm e}^{ikz_g}+f^{\rm even}_{\xi}(k){\rm e}^{ ik|z_{g}|}\right.\nonumber\\
 &  & \left.+f^{\rm odd}_{\xi}(k){\rm sign}(z_g){\rm e}^{ ik|z_{g}|}\right]\phi_{0}(z_{e}).\nonumber\\
\end{eqnarray}
Here $f^{\rm even}_{\xi}(k)$ and $f^{\rm odd}_{\xi}(k)$ are the effective 1D scattering amplitudes for the even and odd partial waves, respectively, and they are given by
\begin{align}
\label{eqn:1df}
f^{\ell}_{\xi}(k)=-i\sqrt{2\pi}\frac{m_g}{\hbar^{2}}\frac{g_{\xi}}{k}\int dz'F_{\ell}(kz')\phi_{0}^{*}(z')\psi^{+}_{k}(z',z'),
\end{align}
for $\ell=({\rm even,\ odd})$, respectively, with $F_{\rm even}(kz')=\cos(kz')$ and $F_{\rm odd}(z')=-i\sin(kz')$. Here the function $\psi^{+}_{k}(z',z')$ satisfies the integral equation
\begin{align}
\label{eqn:ie}
\psi_{k}^{+}(z',z')=\psi_{k}^{0}(z',z')+&g_{\xi}\int dz''G_{0}(z',z';z'',z'')\times\nonumber\\
&\hspace{2cm}\psi_{k}^{+}(z'',z''),
\end{align}
which is a straightforward result of Eq.~(\ref{eqn:scattering-state}).

In this paper we focus on the low-energy limit $k\rightarrow 0$. In this limit 
the scattering amplitude for the odd partial wave is negligible and
the scattering of even partial wave is described by the effective one-dimensional scattering length $a^{\rm eff}_{\xi}$ which is defined as
 \begin{align}
 \label{aeff}
a^{\rm eff}_{\xi}=\lim_{k\rightarrow 0}\frac{i}{k}\left[1+\frac{1}{f_{\xi}^{\rm even}(k)}\right].
\end{align} 
Explicitly, for $k\rightarrow 0$ we have
\begin{align}
f^{\rm even }_{\xi}(k)\approx-\frac{1}{1+ika^{\rm eff}_{\xi}},\quad\  f^{\rm odd }_{\xi}(k)\approx 0.
\end{align}
This scattering amplitude can be reproduced by an effective one-dimensional delta potential 
\begin{align}
\label{geff}
g_{\xi}^{\rm eff}\delta(z_g)\equiv-\frac{\hbar^2}{m_ga^{\rm eff}_{\xi}}\delta(z_g).
\end{align}
Thus, at the low-energy the itinerant $|g\rangle$-atoms experience an effective potential $g_{\xi}^{\rm eff}\delta(z_g)$.
Hence, the place where $g_{\xi}^{\rm eff}$ diverges (i.e. $a^{\rm eff}_{\xi}=0$) is the scattering resonance in the channel $\xi$. 

By scale $z_g\rightarrow z_g/b$, $k\rightarrow kb$, and $\phi\rightarrow \phi\sqrt{b}$, one can show that Eqs.~(\ref{eqn:scattering-state}, \ref{eqn:1df}, \ref{eqn:ie}) become dimensionless, in which the only parameters are the mass ratio $m_g/m$ and the dimensionless interaction strength $\tilde{g}_\xi$ defined as  
\begin{align}
\label{eqn:tildeg}
\tilde{g}_\xi\equiv\frac{g_{\xi}mb}{\hbar^{2}}.
\end{align}
The strength of effective potential $g^{\rm eff}_\xi$ also becomes dimensionless, which is defined as 
\begin{align}
\label{eqn:tildegeff}
\tilde{g}_\xi^{\rm eff}\equiv\frac{g_{\xi}^{\rm eff}m_{g}b}{\hbar^{2}}=-\frac{b}{a^{\rm eff}_{\xi}}.
\end{align}
$\tilde{g}_\xi^{\rm eff}$ and $a^{\rm eff}_{\xi}$ are universal functions of $m_g/m$ and $\tilde{g}_\xi$.

In our calculation, we first numerically solve Eq.~(\ref{eqn:ie}) to derive the function $\psi_{k}^{+}(z',z')$. Substituting the solution into Eq.~(\ref{eqn:1df}) and (\ref{aeff}), 
we obtain the one-dimensional scattering length $a^{\rm eff}_{\xi}$.
Using this result and Eqs.~(\ref{geff}, \ref{eqn:tildeg}, \ref{eqn:tildegeff})
we finally derive the universal relation between $\tilde{g}_\xi^{\rm eff}$ and $\tilde{g}_\xi$, $m_g/m$.

\begin{figure}
\begin{center}
\includegraphics[width=2.9in]{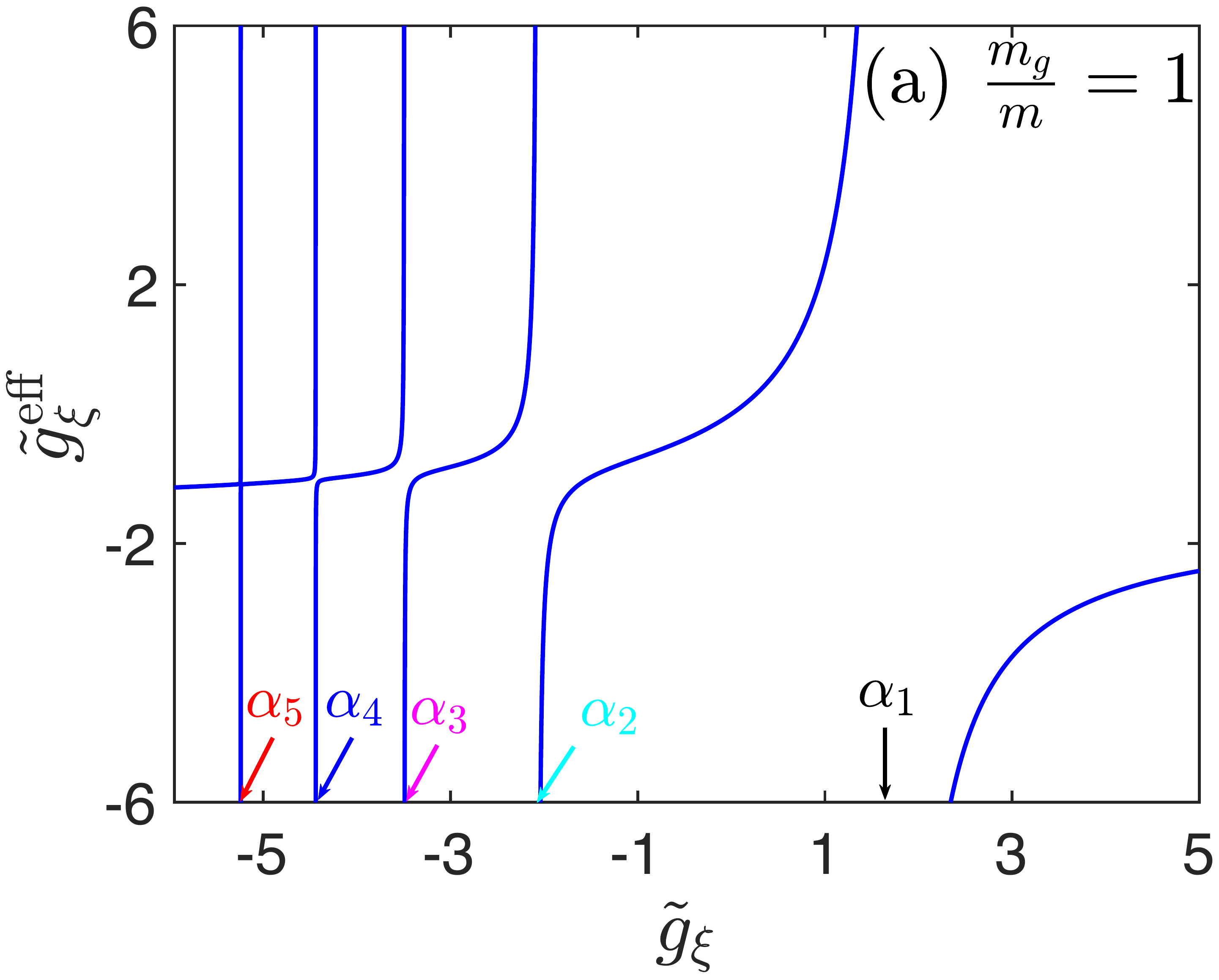}
\includegraphics[width=2.9in]{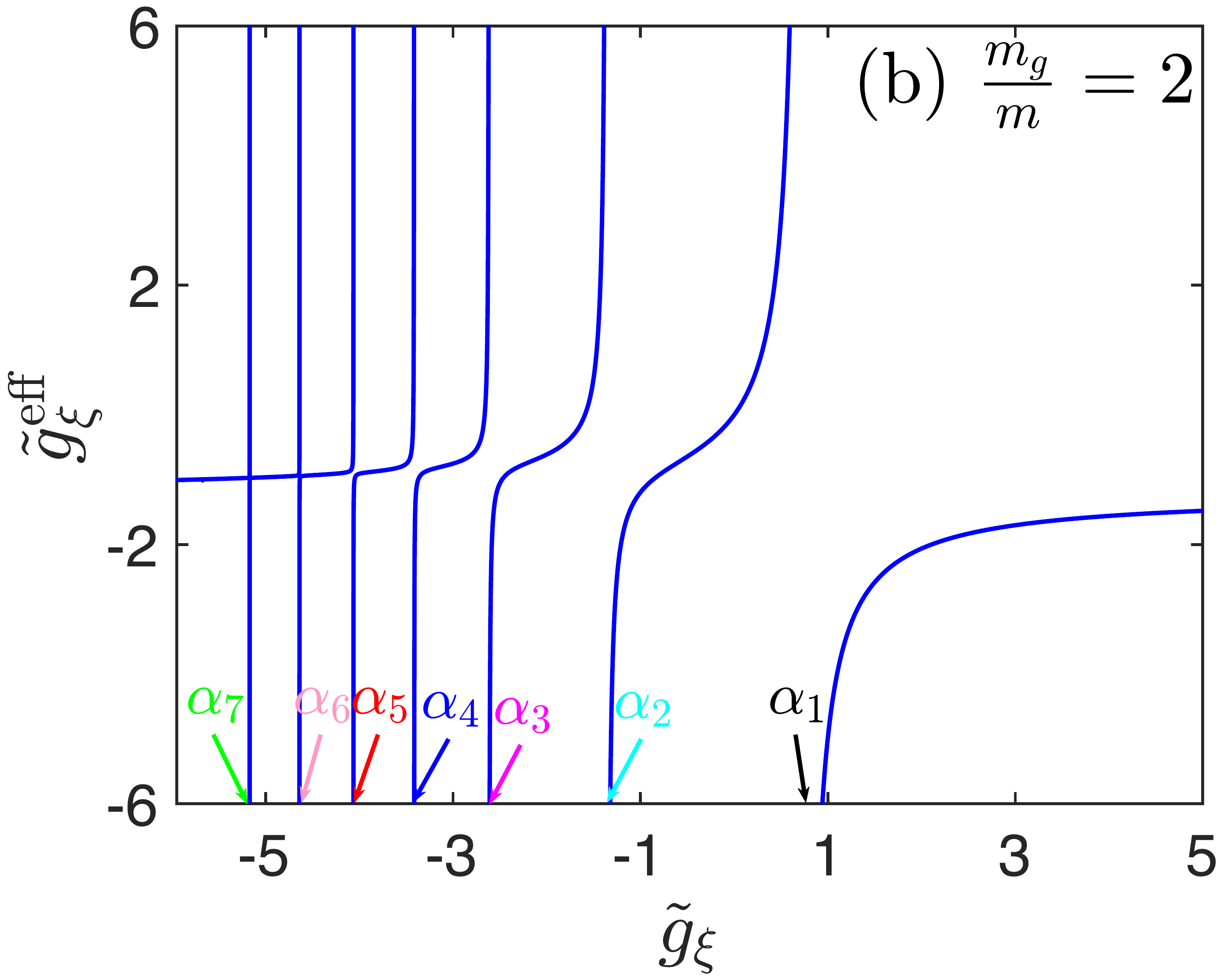}
\end{center}
\caption{ The dimensionless strength for the effective potential $\tilde{g}_\xi^{\rm eff}$ as a universal function of 
the dimensionless bare interaction strength $\tilde{g}_\xi$ with $m_g/m=1$ (a) and $m_g/m=2$ (b).
\label{fig:g-a1d}}
\end{figure}

\section{Results and analysis}
In Fig.~\ref{fig:g-a1d} we show $\tilde{g}_\xi^{\rm eff}$ as a universal function of $\tilde{g}_\xi$, for the mass ratio $m_g/m=1$, $m_g/m=2$, respectively. Here we can see multiple resonances, and the $n$th of them occurs at $\tilde{g}_\xi=\alpha_n$. For instance, for $m_g/m=1$ we have $\alpha_1=1.71$, $\alpha_2=-2.07$, $\alpha_3=-3.49$, $\cdots$, while for $m_g/m=2$,  we have $\alpha_1=0.73$, $\alpha_2=-1.36$, $\alpha_3=-2.62$, $\cdots$.  Similar phenomenon has been obtained in other mixed dimension systems \cite{yvan-mixd-d,tan-mixd-d}.

The occurrence of a series of resonances with negative $\tilde{g}_\xi$ can be understood as follows. Our problem is essentially an infinite-channel scattering problem, and the $n$th channel corresponds to the $|e\rangle$-atom being in the $n$th eigen-state of the axial harmonic trap. The threshold energy for the $n$th channel is $n\hbar\omega_z$ (here we have ignored the zero-point energy of the harmonic trap). Thus, when the kinetic energy of the incoming particle $\hbar^2k^2/(2m_g)<\hbar\omega_z$, only the zeroth channel is open, and all the other channels are closed. For $\tilde{g}_\xi<0$, this interaction potential can support one bound state in each closed channel. When the energy of this bound matches the threshold of the open channel, a resonance occurs. Therefore, each closed channel will give one resonance in the negative $\tilde{g}_\xi$ side. The higher the resonance, the smaller the wave function overlap, and consequently, the narrower the resonance width.

\begin{figure}[t]
\begin{center}
\includegraphics[width=3.2in]{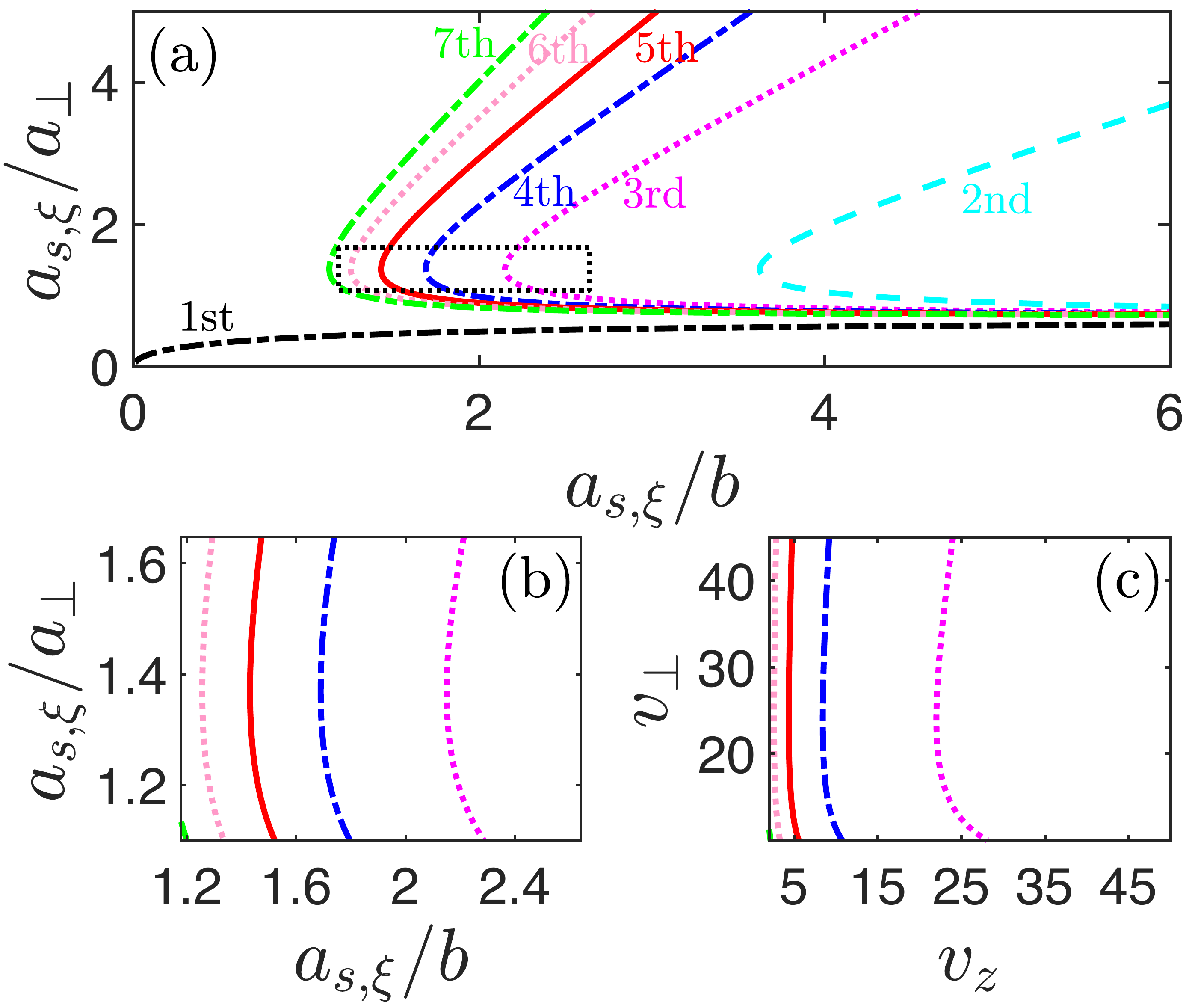}
\end{center}
\caption{(a) The position for resonant spin-exchanging interaction in the $a_{s,\xi}/a_{\perp}-a_{s,\xi}/b$ plane for $m_g=m$. The black dotted rectangle indicates the regime that can be attainable with a set of typical parameters for ${}^{173}$Yb atom. (b) The zoom-in plot of (a) for the regime inside the black dotted rectangle; (c) The same plot as (b) with the axes $a_{s,\xi}/a_{\perp}$ and $a_{s,\xi}/b$ converted into the optical lattice depth $v_\perp$ and $v_z$, when the confinement is provided by optical lattice potentials. $v_\perp$ is the transverse lattice depth for both $|g\rangle$ and $|e\rangle$ atoms; and $v_z$ is the axial lattice depth for $|e\rangle$ atoms. In (c), we choose $\xi=+$ and $a_{s,+}=2000a_{0}$ with $a_{0}$ Bohr's radius.
\label{fig:contour1}}
\end{figure}

Note that the effective spin exchanging interaction is proportional to the difference between $g_+^{\rm eff}$ and $g_-^{\rm eff}$, as shown in Eq.~(\ref{eqn:potential}). Hence, the place where one of the $g_{\xi}^{\rm eff}$ ($\xi=\pm$) diverges (i.e., the place where $\tilde{g}_{\xi}=\alpha_n$ $(n=1,2,3,\cdots)$) gives rise to a resonant spin-exchanging scattering. Combining Eq.~(\ref{CIR})  and Eq.~(\ref{eqn:tildeg},) $\tilde{g}_{\xi}$ can be expressed in term of two physical tunable parameters $a_{s,\xi}/a_\perp$ and $b/a_\perp$ as
\begin{align}
\tilde{g}_\xi=\frac{4a_{s,\xi}b}{a_{\perp}^{2}}\left(1-{\cal C}\frac{a_{s,\xi}}{a_{\perp}}\right)^{-1}, 
\end{align}
the resonance position in $a_{s,\xi}/b-a_{s,\xi}/a_\perp$ plane is determined by the following identity
\begin{align}
\label{rp}
4 \frac{\left(\frac{a_{s,\xi}}{a_\perp}\right)^2}
{\frac{a_{s,\xi}}{b}\left(1-\mathcal{C}\frac{a_{s,\xi}}{a_\perp}\right)}=\alpha_n ,\quad n=1,2,3\cdots.
\end{align}

\begin{figure}[t]
\begin{center}
\includegraphics[width=3.2in]{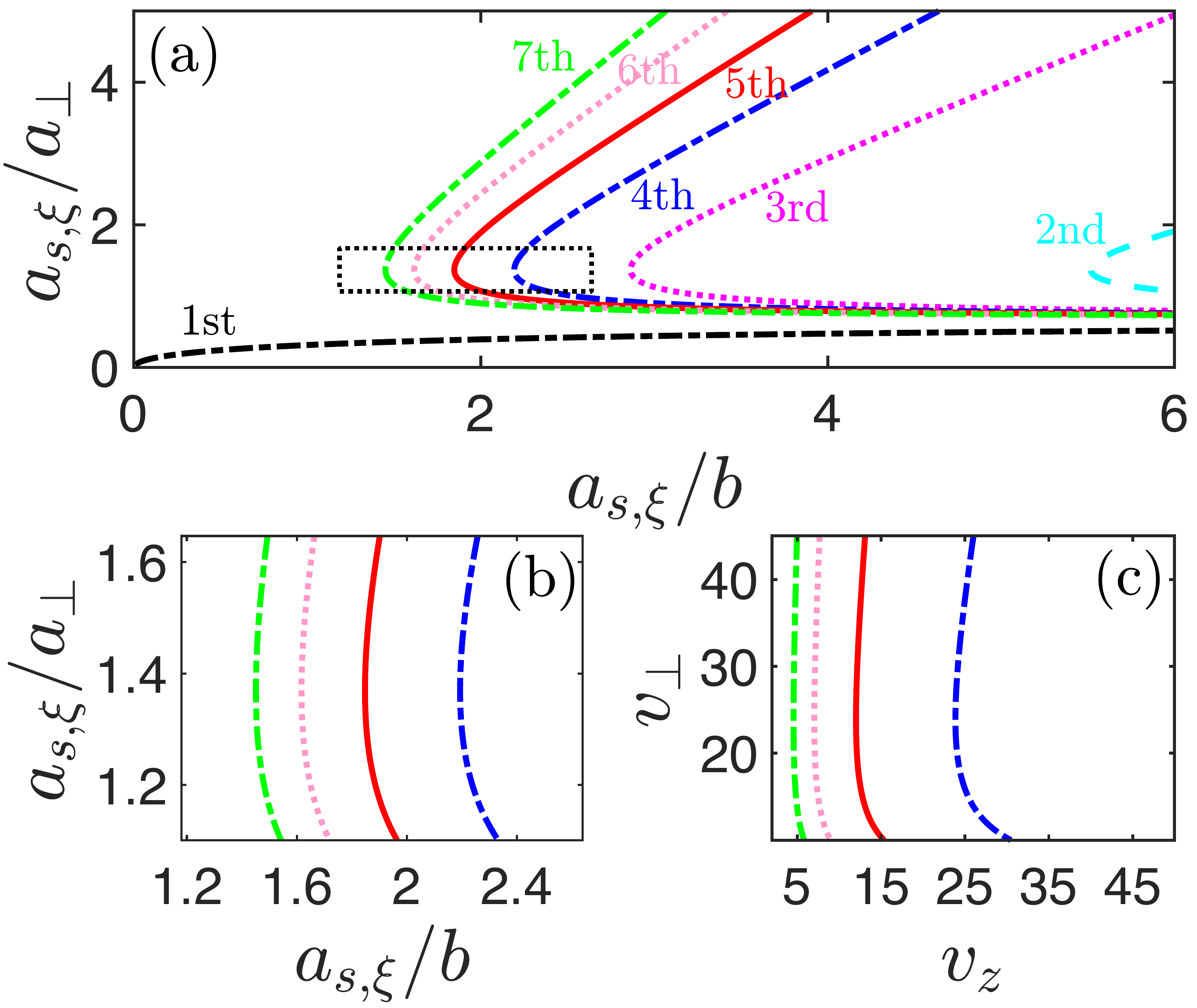}
\end{center}
\caption{The same plot as Fig. \ref{fig:contour1} except for $m_g=2m$. We plot the resonances up to the $7$th one and other more narrower ones are not shown. \label{fig:contour2}}
\end{figure}

In Fig.~\ref{fig:contour1}(a) we show the resonance positions given by Eq. (\ref{rp}) as physical parameters $a_{s,\xi}/b$ and $a_{s,\xi}/a_\perp$ for $m_g=m$. We notice that in the limit of $a_{s,\xi}/b\to\infty$, all the resonance positions will converge to $a_{\perp}/a_{s,\xi}={\cal C}$. That is because, in this limit, the energy difference of the harmonic oscillator is much larger than the kinetic energy, so that the lowest Wannier wave function approximation in Ref.~\cite{ren-kondo} is valid, and the resonance is precisely where $\tilde{g}_\xi$ diverges. 

In practices, the confinements are provided by an optical lattice. Considering an optical lattice of the form $v_\perp E^{\perp}_{\rm R}\left[\cos^2(k_{\perp,0} x)+\cos^2(k_{\perp,0} y)\right]$ for both states, with $E^\perp_{\rm R}=\hbar^2 k^2_{\perp,0}/(2m)$, and an optical lattice of the form $v_z E^z_{\rm R}\cos^2(k_{z,0} z)$ for the $|e\rangle$-state, with $E^z_{\rm R}=\hbar^2 k^2_{z,0}/(2m)$, we have $a_\perp=\lambda_\perp/\left(\sqrt{2}\pi v_{\perp}^{1/4}\right)$ and $b=\lambda_z/\left(2\pi v_{z}^{1/4}\right)$, where $\lambda_\perp=2\pi/k_{\perp,0}$ and $\lambda_z=2\pi/k_{z,0}$. Considering ${}^{173}$Yb atom as an example, $a_{s,+}\approx 2000a_0$, and we use the magic wave length for the transverse direction with $\lambda_\perp=759{\rm nm}$, and for the axial lattice we choice $\lambda_z=670{\rm nm}$ where the ac polarizations of both states differ a lot. 

With the parameters shown above, the ranges of $a_{s,+}/a_\perp$ and $a_{s,+}/b$ can be reached with the tunability of lattice depth are indicated by  the black dotted rectangle in Fig.~\ref{fig:contour1}(a), where one can see that several resonances fall into the rectangle. The zoom-in plot inside the box is shown in Fig. ~\ref{fig:contour1}(b). Then, with these parameters, we can convert $a_{s,+}/a_\perp$ and $a_{s,+}/b$ into $v_\perp$ and $v_z$. The resonance locations as a function of $v_\perp$ and $v_z$ is shown in Fig. ~\ref{fig:contour1}(c). 

Fig. \ref{fig:contour2} shows the same plot as Fig. \ref{fig:contour1}, except for the mass ratio is $m_g=2m$. This is a typical effective mass for a lattice depth of about $5E_\text{R}$. One can see that more higher resonances move into the box. Here we only show up to the $7$th resonance as the higher one will get more and more narrower. Nevertheless, we should also remark that, although these resonances are already quite narrow in the $a_{s,+}/a_\perp-a_{s,+}/b$ plot, it can be reasonably wide in the $v_\perp-v_z$ plot as $a_\perp$ and $b$ are not quite sensitive to $v_\perp$ and $v_z$.

\section{Summary and Outlook}

In summary, we have considered a scattering problem for alkaline-earth atoms in $1+0$-dimensional geometry where the ground state atom is mobile along the tube, while the clock state atom is confined by an axial harmonic trap, with both atoms subjected to a transverse confinement. We have shown that by tuning both the transverse and the axial confinements, and using the realistic parameter for ${}^{173}$Yb as an example, confinement-induced resonance can be reached with practical parameters, where the nuclear spin-exchanging interaction can be enhanced significantly. This will lead to a Kondo effect at relatively higher temperature that can be reached in current experiment. Further work along this line will discuss how to detect the Kondo effect in this system and how to utilize this system to enrich our physical understanding of the Kondo effect. 

\textit{Acknowledgement.} We acknowledge Simon F\"olling for stimulating discussion. This work is supported by NSFC Grant No. 11325418 (HZ), No. 11434011(PZ), No. 11674393 (PZ) and by Tsinghua University Initiative Scientific Research Program (HZ), NKBRSF of China under Grant No. 2016YFA0301600 (HZ), as well as the Research Funds of Renmin University of China under Grant No. 16XNLQ03(PZ).

\end{document}